\documentclass[twocolumn]{aastex631}
\usepackage{natbib}
\usepackage{graphicx}
\usepackage{mathrsfs}
\usepackage{amsmath}
\usepackage{tikz}
\usepackage{hyperref}

\tikzstyle{blue} = [rectangle, rounded corners, minimum width=3cm, minimum height=1cm,text centered, draw=black, fill=blue!30]
\tikzstyle{red} = [rectangle, rounded corners, minimum width=3cm, minimum height=1cm,text centered, draw=black, fill=red!30]
\tikzstyle{green} = [rectangle, rounded corners, minimum width=1.5cm, minimum height=1cm,text centered, draw=black, fill=green!30]
\tikzstyle{orange} = [rectangle, rounded corners, minimum width=1.5cm, minimum height=1cm,text centered, draw=black, fill=orange!30]
\tikzstyle{arrow} = [thick,->,>=stealth]

\begin{document}

\title{Tuning the Rate of Tightly Packed Systems To Produce Planet Occurrence\\ Trends with Galactic Height}

\author{Sarah Ballard}
\affiliation{Department of Astronomy, University of Florida, Gainesville, FL 32611, USA}

\begin{abstract}
The formation of planetary systems has historically been considered in isolation, decoupled from processes on galactic scales. Recent findings employing data from ESA's \textit{Gaia} mission challenge this narrative, identifying trends in planet occurrence with galactic kinematics and stellar age. The findings indicate changes in planet occurrence over and above the predicted changes from metallicity variation within the Milky Way, so that changes to stellar metallicity alone (long understood to be deterministic in planet outcomes) cannot explain the trends entirely. The scope of potential factors influencing planet formation has grown progressively wider, with accompanying theoretical support for galactic-scale influences upon planet formation. In this manuscript, we investigate specifically how changes to the rate of Systems of Tightly-packed Inner Planets (STIPs) could manifest as a trend in planet occurrence with galactic height. We focus our study upon M dwarf planetary systems for two reasons: first, they host STIPs at high rates, and secondly, their longevity makes them useful probes for kinematic trends over Gyr. We consider two models for a varying STIP rate: one in which STIP likelihood is determined by stellar age alone, irrespective of galactic time, and another in which the STIP likelihood suddenly increased in recent galactic history. Both models, which impose a higher STIP likelihood among younger stars, produce a negative gradient in planet occurrence with increasing height from the galactic midplane. We find that a step function model in which STIP likelihood increased by a factor of several $\sim$a few Gyr ago resembles an observed trend among FGK dwarfs. We consider plausible physical mechanisms that could mimic the hypothesized model, given known links between STIP occurrence and other stellar and planetary properties
\end{abstract}


\section{Introduction}

 The \textit{Gaia} data releases \citep{prusti_gaia_2016, vallenari_gaia_2023} have revolutionized our understanding of galactic dynamics, allowing us to situate stellar and planetary astronomy within a broader galactic context. Historically, the process of planet formation has been considered to occur in an isolated star-disk system \citep{armitage_dynamics_2011, williams_protoplanetary_2011, winn_occurrence_2015}. Interaction between an individual protoplanetary disk and its star's immediate natal environment, via evaporative winds or dynamical interactions \citep{adams_photoevaporation_2004, cai_signatures_2018, de_juan_ovelar_can_2012, ansdell_alma_2017}, extend the story's scope. More recently, evidence has emerged that clustering in the stellar birth environment affects planet outcomes. \cite{winter_stellar_2020} and \cite{kruijssen_bridging_2020} leveraged \textit{Gaia} measurements to demonstrate that planet occurrence changes in ``high-density" versus ``low-density" stellar phase space. In ``high-density" phase space, planet host stars share a greater number of co-moving stars than unstructured space. \cite{hamer_hot_2019} also identified a link between planet occurrence (specifically of Hot Jupiters) and stellar kinematics. These findings indicate that the star-planet system is not isolated; its context extends \textit{at least} as far as neighboring stars in the same star-forming region. 
 
 The scope has grown progressively wider. \cite{winter_prevalent_2020} posited a specifically \textit{galactic} framework for planet formation, in which ``local" conditions matter on much larger scales. Galactic-scale properties of the interstellar medium, they argued, as well as the proximity of massive stars to drive far-ultraviolet mass loss, shape protoplanetary disk outcomes. Both \cite{bashi_small_2019} and \cite{dai_planet_2021} identified trends between the relative galactic velocities of stellar hosts and their planetary populations. This story, however, is made more complex by the links between galactic motion and metal content of stars \citep{gandhi_high-_2019}. Planets from different galactic populations (i.e. belonging to the thick disk or thin disk) are also marked by differences in abundance, both in iron and $\alpha$ elements \citep{santos_constraining_2017}. One possibility is that the observed effect of galactic kinematics upon planet occurrence is incidental, with isolated stellar metallicity being the actual determinant of planet outcomes. In this sense, galactic trends in planet occurrence might be mere proxies for the metal content of the protoplanetary disk. On the other hand, larger-scale processes may be deterministic in ways previously unconsidered within the planet-formation-in-isolation picture. For example, \cite{winter_planet_2024} proposed that galactic-scale turbulence may modulate the late-stage infall of material from the interstellar medium onto the protoplanetary disk. 
 
 A recent breakthrough finding disentangled, at least in part, the effects of galactic motion versus metallicity. \cite{zink_scaling_2023}, employing a sample of stars observed by the \textit{Kepler} spacecraft \citep{Borucki10}, found that small planet occurrence around FGK dwarfs is correlated with height above the galactic midplane. Intriguingly, this relationship is stronger than what would be expected from the galactic metallicity gradient alone. This is suggestive, given that the motions of stars within the galaxy encode age information: generally speaking, older stars exhibit higher velocity dispersion or vertical action as a population \citep{wielen_diffusion_1977, rocha-pinto_chemical_2004, almeida-fernandes_method_2018}. The fact that the trend in galactic height with metallicity does not, by itself, trace the change in planet occurrence may indicate an age-dependent process at work.  
 
 Our aim in this manuscript is to approximate how a planet occurrence model that changes in time would manifest as a function of height from the galactic midplane. To this end, we set out to construct a synthetic catalog of stars over a realistic range of galactic height, populate those stars with planets according to a prescription, ``observe" its planets in slices of galactic height, and then compare them to the real observations. Such a catalog, in order to be useful, must at least roughly approximate the true age/galactic height relationship of real Milky Way stars. For the purposes of this manuscript, we employ highly simplified toy models for the star formation history (thus creating a realistic sample of stellar ages) and the age/galactic height dispersion relationship of the Milky Way galaxy. With these models, albeit simplified, we can in theory trace an age-dependent planet phenomena to its corresponding presentation with galactic height. 
 
 We consider here a very specific type of change to planet occurrence, which is the rate of ``Systems of Tightly-Packed Inner Planets" (STIPs) (see review in \citealt{ford_architectures_2014}). This is an appealing potential mechanism to apply to a scenario in which kinematics plays a large role. Their presence may reflect the local stellar environment at birth, in a way that allows for a possible connection to a galactic context \citep{winter_stellar_2020}. There already exists evidence, described by \cite{miyazaki_evidence_2023} that the rate of Hot Jupiters has changed over galactic time in ways not explainable by metallicity alone, in addition to which \cite{yang_planets_2023} found that planet multiplicity increases around younger stars, controlling for metallicity. The presence of STIPs is itself suggestively linked to the presence or absence of Jovian planets. \cite{Becker20_USP} found that particularly close-in sets of multi-planet systems are often accompanied by a larger outer planets in a  discernibly distinct dynamical state. Hot Jupiters have traditionally been considered solitary objects, in that the existence of a Hot Jupiter often precludes the existence of a close-in set of small planets \citep{wright_ten_2009, steffen_kepler_2012, huang_warm_2016, hord_uniform_2021, wang_transiting_2021, ivshina_tess_2022} (though more recent evidence has revealed that this trend is not universal, per e.g. \citealt{Becker15, wu_evidence_2023}). Several mechanisms have been suggested for why this is so: (1) STIPs are vulnerable to dynamical instability in the presence of a Jovian planet \citep{Becker17, hansen_perturbation_2017, granados_contreras_dynamics_2018}. This instability is stronger as the Jupiter's eccentricity increases \citep{Denham19}, so that the high-eccentricity migration pathway for making Hot Jupiters \citep{rasio_dynamical_1996, Chatterjee08, weidenschilling_gravitational_1996, wu_planet_2003, Fabrycky07b} makes STIP survival unlikely. In the context of the nearby stellar environment, the underlying mechanism to excite high eccentricity could be stellar flybys of planetary systems \citep{wang_hot_2020, wang_transiting_2021, rodet_correlation_2021}. On the other hand, STIPs may actually be \textit{positively} linked to the presence of Hot Jupiters, if (2) the in-situ formation of hot Jupiters is an extreme outcome of early metastability of STIPs \citep{boley_situ_2016}. \cite{wu_evidence_2023} also set forth a framework in which short-period giant planets result from dynamical sculpting in compact multiplanet systems. It's also possible (3) that the small planets in STIPs survive to the present day when migration trapping prevents larger planets from migrating inward \citep{zawadzki_migration_2022}. 
 
 Historically speaking, low-mass dwarfs have been especially useful in calibrating stellar age and galactic kinematics, by virtue of their long lifetimes \citep{reid_palomarmsu_1995, faherty_brown_2009, kiman_exploring_2019, angus_exploring_2020, popinchalk_evaluating_2021}. While the original \cite{zink_scaling_2023} findings focused upon FGK dwarfs from the \textit{Kepler} and K2 missions, we elect to focus this study upon low-mass dwarfs ($T_{\textrm{eff}}<4000$ K, $M_{\star}<0.5M_{\odot}$) for several reasons. Firstly, M dwarfs are ``models of persistence" \citep{Shields16} with lifetimes longer than the lifetime of the Milky Way. For this reason, we can consider their star-and-planet-formation history without having to model the rate at which they then leave the Main Sequence, in the generation of a synthetic sample. Secondly, by modeling planet occurrence among M dwarfs, we are characterizing the vast majority of the galaxy's planetary systems. Planets smaller than Neptune are 3.5 times more abundant around M dwarfs \citep{Mulders15}, and small stars themselves comprise 75\% of stars in the Galaxy \citep{Henry04}. Thirdly, M dwarfs host STIPs at a higher rate than FGK dwarfs: $\sim$20\% (e.g. \citealt{Ballard16, Muirhead15} versus $\sim$5-10\% \citep{lissauer_closely_2011, volk_consolidating_2015, lam_ages_2024}. In this sense, we are attempting to model a dominant trend upon planetary systems in the Milky Way. 

 This manuscript is organized as follows. First, we describe our methodology for generating synthetic samples of stars and planets in Section \ref{sec:Methods}. We employ two prescriptions for varying planet occurrence, and specifically the rate of closely packed planetary systems, with time. We then ``observe" these synthetic planetary systems, gathering 50 samples of transiting planets at a range of galactic heights. We tune the size of this sample to approximate the precision of the data in \cite{zink_scaling_2023}. In Section \ref{sec:analysis} we reverse-engineer the underlying planet occurrence with galactic height from these ``observed" samples. We demonstrate first that we correctly recover the injected ground truth within the predicted confidence intervals. We then fit a power law to these ``observations" of planet occurrence versus galactic height, using the same methodology as \cite{zink_scaling_2023}. In this way, we assess how a varying STIP rate over galactic time would appear as a trend in planet occurrence with increasing distance from the galactic midplane. In Section \ref{sec:results}, we consider our findings in light of possible physical mechanisms before concluding.    

\section{Methods}
\label{sec:Methods}

In order to simulate planet occurrence as a function of galactic height, assuming that STIP fraction changes over time, we require three separate models. Firstly, we need a reasonable synthetic sample of host stars: we describe this process in Section \ref{sec:stellar}. We aim to eventually paint onto this sample of stars a relationship specifically linking stellar age to height above the galactic midplane, described in in Section \ref{sec:galheight}. Secondly, we require a blueprint for the generation of individual planetary systems, which we describe in Section \ref{subsection:planets}. An acceptable blueprint must, at a minimum, successfully reproduce basic statistics from transit surveys: number of planets per star, transit multiplicity, etc. Thirdly, we require a model to describe how that individual planetary blueprint may change with time, and make sure to assign planetary systems to the stellar sample according to that prescription. We consider two toy models in this manuscript, detailed in Section \ref{subsection:decay}, both of which cause the rate of specifically compact multiple planetary systems to vary with galactic time. We generate 25 stellar samples according to the machinery described below in Section \ref{sec:stellar}, going on to apply the first and then the second model prescription for varying planetary occurrence with time among these stars. This ultimately results in 50 synthetic samples of stars and planets, 25 reflective of one time-dependent occurrence model, and 25 reflective of the other. 

\subsection{Generation of Stellar Population}
\label{sec:stellar}

\subsubsection{Stellar properties}

We make a set of several simplifying assumptions, in the generation of the stellar sample. They are as follows:

\begin{itemize}
    \item We assume that the sample of stars is drawn uniformly in linear age, from 0 to 12 Gyr old. 
    \item We assume a fiducial M dwarf host with $T_{\textrm{eff}}=3400$ K. While this is our mean effective temperature, we draw a normal distribution of stellar effective temperatures with a standard deviation of 200 K around this mean. From effective temperature, we then assign each star a radius from \cite{Boyajian12}. 
    \item We assume a common STIP ratio across all effective temperatures: e.g. if 20\% of 4000 K stars host a compact multi, than so too do 20\% of 3200 K stars in our sample.  
\end{itemize}
 
We justify these assumptions for the sake of a relatively simplified experiment, given competing considerations for additional complexity. The generation of a realistic sample of host star ages over galactic time requires a prescription for the star formation history (SFH) in the Milky Way. This is a complex undertaking: star formation in the universe as a whole peaked at the ``cosmic dawn" between redshifts $z=2-3$ \citep{madau_cosmic_2014, lilly_canada-france_1996, madau_star_1998, hopkins_normalization_2006}, and some studies indeed deduce an early star formation peak for the Milky Way (e.g. \citealt{nidever_tracing_2014, revaz_computational_2016}). Yet, other studies conclude that star formation in the Milky Way did not trace the cosmic SFH, finding that star formation actually peaked in the most recent 5 Gyr instead \citep{snaith_reconstructing_2015}.  Simulations of nominal ``Milky-Way-type" galaxies \citep{yin_milky_2009, revaz_computational_2016, garrison-kimmel_origin_2018, hopkins_fire-2_2018}, when observed in their present-day $z=0$ states, exhibit diverse morphologies and kinematics. This is reflected in a wide range of star formation histories, affected by the the accretion of other smaller galaxies onto the synthetic ``Milky Way", the degree of rotation support, and the the relative fractions of mass ending up in the disk from the halo, among other factors \citep{garrison-kimmel_origin_2018}. 

\begin{figure}[htp!]
    \centering
    \includegraphics[width=3in]{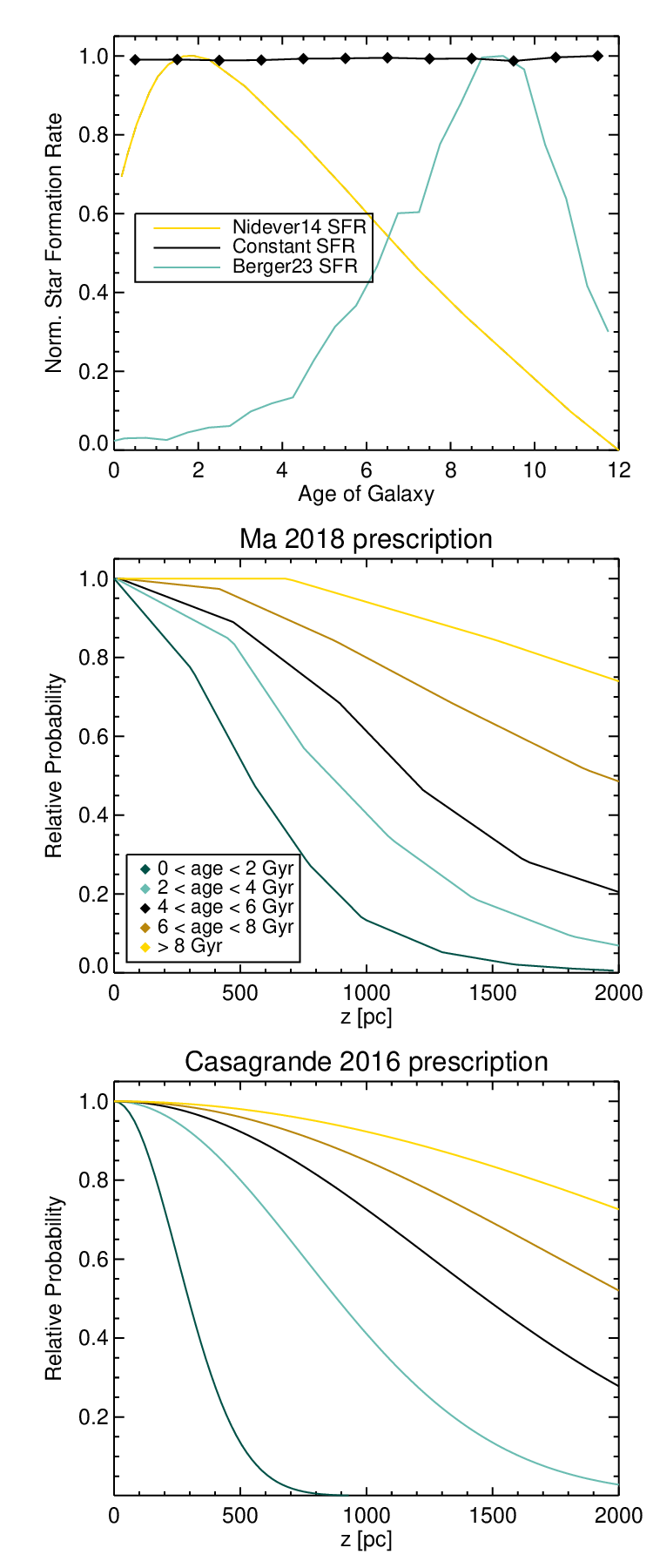}
    \caption{\textit{Top panel:} Examples of star formation history (SFH) models, considered when constructing our stellar sample. While some studies, both of simulations and observations, of the Milky Way's SFH show formation peaking early on (see e.g. semi-analytic model shown here from \cite{nidever_tracing_2014}), the ages of \textit{Kepler} and K2 dwarfs peak much more recently. The curve from \cite{Berger23} corresponds to the inferred time of formation, employing the isochrone ages from that work. \textit{Bottom panels:} Relationships between stellar age and distance from galactic midplane, one from a simulation study \citep{ma_structure_2017}, and one inferred from measurement \citep{casagrande_measuring_2016}.}
    \label{fig:age_midplane}
\end{figure}

However, we must distinguish for the purposes of this study between a truly \textit{representative} distribution of stars drawn from the Milky Way star formation history, and the sample of stars monitored by transit surveys such as \textit{Kepler}, K2, and the Transiting Exoplanet Survey Satellite (\textit{TESS}, \citealt{Ricker14}).  Whether or not the history of the galactic Solar environs, which necessarily comprises most transit survey stars, is itself reflective of the Milky Way's SFH is uncertain \citep{snaith_reconstructing_2015}, though some studies have argued that star formation exhibits homogeneity across the Milky Way \citep{ness_homogeneity_2022}. Using the ``cosmic dawn" SFH, a truly ``representative" age distribution of the galaxy's M dwarfs ought to peak at ages between 8 and 10 Gyr ago; yet the distribution of \textit{surveyed} stars from \textit{Kepler} peak at ages of 2-3 Gyr \citep{Berger20}.  By virtue of observing footprints further out of the galactic plane, K2 included more old stars. Roughly half are older than 6 Gyr \citep{Berger23}, but the \textit{Kepler} sample dominates in the contribution to the planet yield. This is reinforced by the range of galactic heights of the stars included in the \cite{zink_scaling_2023} survey. The \textit{Kepler} stars peak at 290 pc out of the midplane, while the K2 stars peak at 410 pc out of the midplane. At a height of $\sim$ 500 pc, simulations of the Milky Way indicate that 80\% of the stellar mass belongs to the ``thin disk", which is almost entirely comprised of stars younger than 4 Gyr \citep{ma_structure_2017}. Yet, kinematic ages also indicate that the M dwarfs in the \textit{Kepler} sample are older on average then the F and G spectral type stars. We opt for a simplifying assumption of a flat age distribution between 0 and 12 Gyr. We show a comparison between a fiducial star formation rate in \cite{nidever_tracing_2014}, the rate of star formation inferred from the ages of \textit{Kepler} and K2 planet hosts from \cite{berger_gaia-kepler-tess-host_2023}, and our adopted uniform occurrence rate in Figure \ref{fig:age_midplane}.  

We have also made a simplying assumption in the choice of a fiducial mass for a star in the sample. An experiment designed to model the true galactic trend would reflect the actual distribution of M dwarfs in nature: they are drawn from a bottom-heavy IMF that depends upon both metallicity and stellar age \citep{li_stellar_2023}. A survey sample that reflects the true population ought therefore to peak at the lowest mass M dwarfs. However, a comparison to real transit surveys would reflect the opposite, and peak at \textit{higher} mass M dwarfs. \textit{Kepler}'s $\sim$5000 surveyed M dwarfs \citep{Brown11, Dressing13, Dressing15} peaked at 0.5 $M_{\odot}$, with only 10\% later than spectral type M3V \citep{Muirhead15, Hardegree19}. A comparison between the planet hosts detected by \textit{Kepler}, K2, and \textit{TESS} from \cite{berger_gaia-kepler-tess-host_2023} shows that the peak at earlier spectral type becomes less strong: among characterized M dwarf K2 and \textit{TESS} candidates the host stars are flat across mid-to-early M dwarfs, with roughly as many hosts at 3300 K as at 4000 K. By selecting a fiducial M dwarf host star at 3400 K (roughly spectral type M2V, per \citealt{Boyajian12}), we compromise in experimental design between modeling the underlying galactic trend and reflecting the trend as predicted among surveyed exoplanets. 

Finally, we opt for the simplying assumption of a common rate of compact multiple planetary systems across the range of stellar masses in our synthetic sample. In reality, there appears to be some variability in this rate in the M spectral type: for mid-M dwarfs, \cite{Muirhead15} found that the 15\% rate for early-type M dwarfs identified in \citep{Ballard16} increases to 20\% for mid-type M dwarfs.

\subsubsection{Galactic height}
\label{sec:galheight}
Given the synthetic sample of stars described above, we now consider the assignment of galactic height based upon stellar age. As stars age, on average their interaction with massive star-forming regions or the galactic potential increases their distance from the galactic midplane \citep{wielen_diffusion_1977, rocha-pinto_chemical_2004, almeida-fernandes_method_2018}. Observational studies of this relationship have been revolutionized by Gaia \citep{prusti_gaia_2016, vallenari_gaia_2023}, by measuring the kinematic properties of large swaths of the Milky Way's stars \citep{mackereth_dynamical_2019, vieira_vertical_2023}. Given stellar age metrics, the dispersion over time from the galactic midplane can be constrained in practice. For example, the Milky-Way-type galaxy simulation by \cite{carrillo_relationship_2023} exhibited an increasing dispersion from the galactic midplane with age. Stars older than 8 Gyr showed roughly twice the dispersion of stars 4 Gyr old: 1 kpc in standard deviation from the midplane, versus 500 pc. With another simulation of a Milky-Way-type galaxy, \cite{ma_structure_2017} identified the contribution to the stellar population at each galactic height from different age slices, in increments of 2 Gyr: we include these distribututions in Figure \ref{fig:age_midplane}. This is in alignment with the observational findings of \cite{casagrande_measuring_2016}, who derived a relation of $\sim$1 kpc 4 Gyr$^{-1}$. We show in Figure \ref{fig:age_midplane} the predicted resulting midplane heights for different age slices within the sample for two assumptions. One relationship applies a dispersion that is linear in time, normalized to \cite{casagrande_measuring_2016}'s empirical 1 kpc 4 Gyr$^{-1}$ finding. Using this metric, stars with an age of 2 Gyr would be distributed about the galactic midplane with a standard deviation of 500 pc, e.g. The other \ref{fig:age_midplane} relationship shows the distributions about the midplane from \citep{ma_structure_2017}'s simulations of the a Milky-Way type galaxy (we have normalized all distributions to peak at 1 here, for comparison of the spread. The original figure from \cite{ma_structure_2017} is scaled by the stellar mass per unit volume in each slice). The reader can see that these distributions are approximately similar, differing by roughly a factor of two in spread for the same age interval. Ultimately, we elect to employ an assumption of approximate linear increase in dispersion with time, using the dispersion rate of \cite{casagrande_measuring_2016}. 

\subsubsection{Sample size}
\label{sec:sample_size}

Ultimately, we aim to extract synthetic planet occurrence ``measurements" from our synthetic sample, for slices in galactic midplane height between 100 and 1000 pc. For illustrative purposes, we aim for approximately the same degree of measurement uncertainty as in \cite{zink_scaling_2023}. In this way, we can mimic an experiment with a sample of real M dwarfs as closely as possible, fitting a power-law function to occurrence versus galactic height with approximately the same precision. We therefore tune our sample size to approximately match the sample size in that work. About 70,000 stars from the combined \textit{Kepler} and \textit{K2} missions contributed to the \cite{zink_scaling_2023} occurrence calculations between 100-1000 pc. We focus here upon the M dwarf population, rather than FGK dwarfs, so that the number of planets per star is between 3--4 times higher \citep{Mulders15}. For this reason, we need only synthesize about 1/3rd the sample size, over the same range in $z$, to expect the same planet yield (and therefore similar measurement uncertainty). However, it is also useful to examine the predicted trend from the midplane out to higher galactic height. For this reason, we employ a similar sample size of 70,000 stars, but distribute them from z=0 out to 3000 pc above and below the midplane according to their ages, as described in more detail below (Section \ref{sec:galheight}). This exercise naturally (and realistically) places the most stars at $<100$ pc from the midplane, but still results in a sample size between 100-1000 pc of 16,000 stars. In practice, this results in a mean number of detected planets of $\sim$1000 orbiting stars with $z$ between 100 and 1000 pc, a comparable figure to the 853 planets contributing to the occurrence calculation over that range in \cite{zink_scaling_2023}. 

\subsection{Generating planetary systems}
\label{subsection:planets}

We generate the set of planet radii and orbital periods from the occurrence rates in \cite{Dressing15}, in the same manner as \cite{Sullivan15}. We employ a fixed resolution in both log(period) and log(radius) (inherited from \citealt{youdin_exoplanet_2011, Howard12, Dressing13}, and others), with an approximate spacing of 1 dex between adjacent log(period) bins and 0.2 dex between adjacent log(radius) bins.  In practice, the index i spans periods from 0.5 to 200 days in 13 regular log intervals of 1 dex, and the index j spans radii from 0.3 to 4$R_{\oplus}$ in 17 regular log intervals of 0.2 dex. 

When drawing from the number of planets per star and their mutual inclination, we employ the mixture model of \cite{Ballard16}. Both in that study and in \cite{Muirhead15}, the authors found that $\sim$20\% of M dwarfs host ``compact multiple" STIPs, with the remaining 80\% hosting planetary systems that are either more sparsely populated, or with higher mutual inclinations. The former sample is defined by $\ge2$ planets interior to 10 days in \cite{Muirhead15}, and by $\ge5$ planets with orbital periods less than 200 days in \cite{Ballard16}. Systems with more than one transiting planet will be drawn preferentially from flatter (low mutual inclinations), compact systems of multiple planets due to higher geometric likelihood. Correspondingly, systems with more than one transiting planet tend to be dynamically cool, using the ``dynamical temperature" framework of \cite{Tremaine15}. The mutual inclination distribution peaks at $<2^{\circ}$, per \cite{fabrycky_architecture_2014} for FGK dwarfs and \citealt{Ballard16} for M dwarfs, with eccentricities $<0.05$ (per \citealt{VanEylen_2019} for FGK dwarfs, \citealt{sagear_orbital_2023} for M dwarfs). Other studies have shown that a model with a smooth range of dynamical temperature does just as well, if not better, than two distinct distributions at reproducing population metrics of transiting planets, such as number of transiting planets per star \citep{Zhu18,Millholland21, He19, He20}. We adopt the two-population model as a heuristic in this work, in order for ease of investigation into how planet occurrence is affected by changing the contribution of one population versus another. In the STIP population, we draw either 5, 6, or 7 planets with equal probability, assigning mutual inclinations randomly from a Rayleigh distribution peaking at 2$^{\circ}$ (determined for the \textit{Kepler} M dwarfs per \citealt{Ballard16}). For the non-STIP population, we draw between 1--3 planets with equal probability, with mutual inclinations assigned from a Rayleigh distribution peaking at $8^{\circ}$. We assign eccentricity per \cite{sagear_orbital_2023}: STIP eccentricities are drawn from a Rayleigh distribution peaking at $e$=0.02, and non-STIPs from a Rayleigh distribution peaking at $e$=0.24. The longitudes of periapse are assign in a random uniform fashion from 0 to 360$^{\circ}$. 

 We assign planetary masses with the relations of \cite{zeng_simple_2017} for planets $<1.5$ $R_{\oplus}$ and \cite{Wolfgang16} for planets $>1.5$ $R_{\oplus}$. We assess the stability of the system by ensuring that planets satisfy the mutual Hill separation criterion defined in \cite{Fabrycky12a}, redrawing new sets of orbital period and radii until the criteria are satisfied. 

 To generate a synthetic ``observed" data set, we randomly assign a midplane to each planetary system. Assuming systems are randomly distributed on the sky, we draw midplanes uniformly in sin$(i)$. While technically planets transit with any impact parameter $b<1$, in practice the sample of detected M dwarf from \textit{Kepler} have impact parameters mostly $<$0.8 \citep{Swift15} rather than uniform in $b$ from 0--1. We therefore assign ``detected" status in a probabilistic manner with impact parameter $b$, applying the empirical relation from \cite{Swift15}, with probability of detection decreasing sharply after 0.8. We then employ the average sensitivity function in \cite{Dressing15} to determine whether a given transiting planet is ``detected" by drawing a random uniform number from 0 to 1. If the detection probability corresponding to that planet's radius and period is higher than the randomly drawn number, we record its true radius and period in the catalog of synthetic detections. 

\subsection{Imposing occurrence trends with time}
\label{subsection:decay}
We aim to investigate how a change in the rate of compact planetary systems with time, over the age of the galaxy, can manifest as an apparent trend in decreasing planet occurrence with galactic height $z$. Given the suggestive trend among FGK dwarfs (an apparent increase in planet occurrence close to the midplane), we favor a model in which \textit{younger} planetary systems are likelier to reside in a compact multiple configuration. We now consider how to assign planetary systems to each star in our sample.  With a toy model in which \textit{younger} stars are likelier to host tightly packed planetary systems, we can expect a pile-up of \textit{apparently} higher planet occurrence among the more youthful midplane stars. To calibrate the approximate behavior of the model, we consider the observed trend by \cite{zink_scaling_2023}: a decrease in planet occurrence by a factor of $\sim2$ as $z$ increases from 100 to 1000 pc from the midplane. Given the typical drift in galactic height of 1 kpc per 4 Gyr \citep{casagrande_measuring_2016}, we conclude a model trend likely needs to manifest over Gyr timescales. Separately, any reasonable model trend must \textit{also} result in the correct $\sim$20\% rate of compact multiples among the galaxy's M dwarfs in the present day \citep{Muirhead15, Ballard16}. 

\begin{figure*}
    \centering
    \includegraphics[trim={0.5cm 0 0.5cm 0}, width=6.0in]{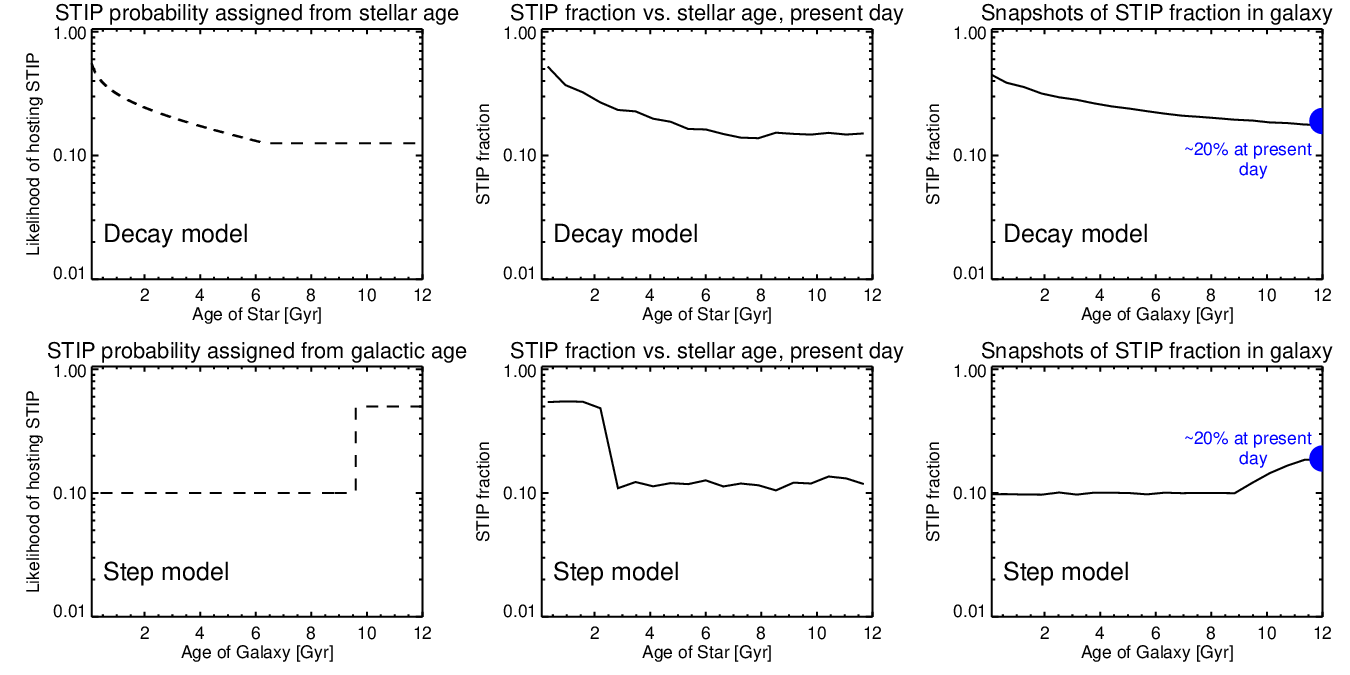}
    \caption{Prescriptions for changing the rate of systems of tightly-packed inner planets (STIPs) with time.  \textit{Top panels:} A ``decay" model that links STIP likelihood to stellar age, applied uniformly over galactic time. \textit{Bottom panels:} A ``step" model that links STIP likelihood to the age of the Milky Way. Both models produce an effect in which STIPs reside around younger stars (see middle panels), though they produce different histories of the STIP rate over the age of the Milky Way before reaching the present-rate rate of $\sim$20\% (right panels).}
    \label{fig:stip_models}
\end{figure*}

We adopt two models by which younger M dwarfs are likelier, on a $\sim$Gyr timescale, to host compact multiple systems of planets. These two scenarios are depicted in Figure \ref{fig:stip_models}. Within the first framework (top panels), we assume that individual planetary systems change with time. We denote this scenario the ``decay" model, by which dynamically packed planetary systems are metastable at birth \citep{Pu15, Volk15, zinzi_anti-correlation_2017}. Over Gyr timescales, the originally densely populated, dynamically cold planetary system is self-disrupted, resulting in less planets, with larger spacings and mutual inclinations. The tendency for younger systems to still be the ``compact multiple" state drives the seeming increase in planet occurrence toward the galactic midplane. In this sense, the probability of a star hosting a compact system of multiple planets is determined from \textit{stellar} age. This is a similar physical scenario to the one proposed by \cite{yang_planets_2023}, in order to explain an apparent increase in number of planets per star and decrease in mutual inclination among younger stars.  In the top panels of Figure \ref{fig:stip_models}, we show the functional form for this probability, applied to each star separately: an individual STIP rate of $\sim$70\% at birth is followed by a fiducial power-law decay in the likelihood of compact multiple status over a period of $\sim$10 Gyr, after which time the likelihood stays constant (that is, we assume that this putative dynamical sculpting, when it occurs, takes place in the first 10 Gyr). This resembles in functional form the decay laws investigated by \cite{lam_ages_2024} and \cite{Teixeira22}. Over galactic time, this results in a decreasing STIP rate in the galaxy as it ages. In the Milky Way's first 2 Gyr, the rate of compact multis is roughly 50\%, as many systems have not yet self-excited. As time passes, fractionally more systems have aged into the dynamically hotter state, arriving by 12 Gyr at the $\sim20$\% fraction of compact multiples that currently characterizes the galaxy's M dwarf planetary systems \citep{Muirhead15, Ballard16}. We depict this evolution in two ways in Figure \ref{fig:stip_models}. For each model, we show both (1) the present-day distribution in STIP fraction as a function of age in the Milky Way. We also show (2) snapshots of how this overall fraction as changed as the galaxy has aged, ending in the $\sim$20\% rate of the present day. 

Within the second framework, planetary systems themselves do not change with time; rather, the fraction of planetary systems born as STIPs increases at some \textit{galactic} time. We designate this the ``step" model, with a fiducial step function increase in STIP likelihood that took place a nominal 2 Gyr ago. Within this framework, stars born in the galaxy's first 10 Gyr have a 10\% likelihood of being STIPs, while stars born in the most recent 2 Gyr had an 50\% likelihood of being STIPs. Just like the ``decay" model, this produces the desired increase in STIP fraction among younger stars when applied to our sample of stellar ages (see middle panels of Figure \ref{fig:stip_models}). However, the story of how the galaxy arrived at this fraction looks very different, viewing snapshots of the galaxy's evolution. Compared to the ``decay" model, this produces the opposite trend (rightmost panels of Figure \ref{fig:stip_models}). For the first 10 Gyr of the Milky Way's lifetime with the ``step" model, the STIP fraction was lower than it is today: stuck at the designated 10\% rate. Only for stars born in the last 2 Gyr does the scenario change: 50\% of these are STIPs. Thus, the rate of compact multiple planetary systems has been increasing for the past 2 Gyr. This is in comparison to the ``decay" model, in which the overall STIP rate across all stars has only been decreasing as the galaxy ages. 

We employ the prescriptions described above, and shown in the leftmost panel of Figure \ref{fig:stip_models}, to assign the likelihood of hosting a STIP to each star. We draw a number from a random uniform distribution between 0 and 1 for each star. If the drawn number is less than the ``STIP probability", then this star is the host to a compact multiple system of planets. If the drawn number is greater, then the star hosts a more sparsely populated and more highly mutually inclined system of planets (described above in Section \ref{subsection:planets}).

\begin{figure*}
    \centering
    \includegraphics[trim={0.5cm 0 0.5cm 0}, width=7.0in]{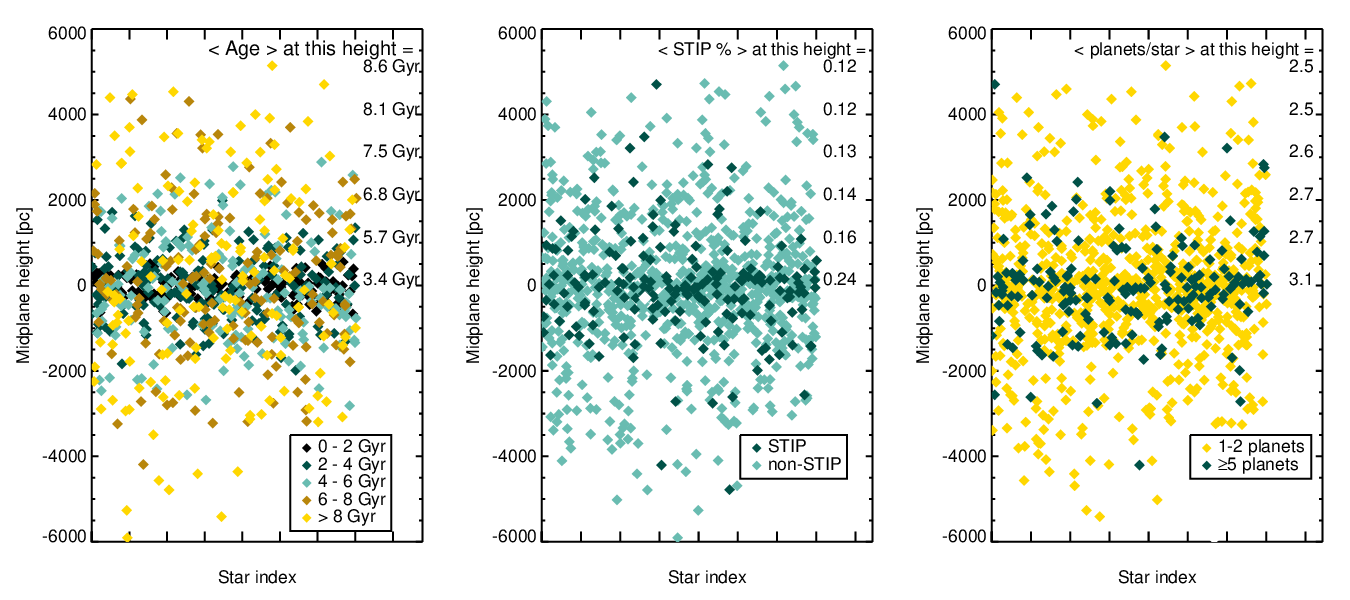}
    \caption{One synthetic sample of star and planetary systems generated per Section \ref{sec:Methods}, shown as a function of height above the galactic midplane. From left to right, color-coded by (1) age, (2) STIP status, and (3) number of planets per star. Averages in these quantities with increasing galactic height are shown along the right of each plot.}
    \label{fig:midplane_sample}
\end{figure*}

We show a representative sample in Figure \ref{fig:midplane_sample}, using the ``step" function for this particular example to generate STIP likelihood. As a function of an arbitrary index number, we show the sample's distribution from the galactic midplane in three ways, after applying the prescriptions described above for generation of the stellar sample, assignation of the STIP likelihood, and generation of planetary systems. First, the sample is color-coded by age; we also overplot the average age at a number of slices at increasing galactic height. Secondly, the sample is color-coded by STIP status, where we overplot the fraction of STIP planetary systems at the same slices in galactic height. Finally, the sample is color-coded by the number of planets per star. The reader can examine visually the variation at $|z|<$1000 pc, the range examined in \cite{zink_scaling_2023}. The average age at these two height slices is 3.4 Gyr and 5.7 years, respectively, resulting in an average age difference of 2.3 Gyr between stars at 0 pc from the midplane and stars at 1000 pc. Over that same range, the fraction of STIPs decreases from 24\% to 16\%. It is useful to compare this fiducial model to the findings of \cite{yang_planets_2023}, who found that the number of planets per star increases around younger stars: while stars younger than 1 Gyr host 1.6-1.7 planet star$^{-1}$, stars with ages $\sim$8 Gyr host 1 planet star$^{-1}$, with 2-3$\sigma$ confidence. This increase by a factor of 1.7 between ``young" and ``old" stars trends similarly with the sample shown in Figure \ref{fig:midplane_sample}, by which the youngest stars in the midplane host 3.1 planets star$^{-1}$, while among stars with average age of 8 Gyr, the planets star$^{-1}$ has fallen to 2.5; this is an increase of a factor of 1.3. 

\section{Analysis}
\label{sec:analysis}

After the generation of the synthetic stellar and planetary samples described in Section \ref{sec:Methods} according to prescriptions for STIP rate over time, we now examine the resulting trends with galactic height. At left in Figure \ref{fig:all_sims} are 25 iterations of the trend in galactic height versus planet occurrence between 100-1000 pc, employing both the ``decay" (top panel) and ``step" (bottom panel) models in STIP rate. We bin the sample at intervals of 250 pc. As predicted, given the imposed condition that younger stars are likelier to be STIPs, planet occurrence is higher toward the more youthful midplane. The overall normalization of the galactic-height-versus-planet occurrence models varies, but the \textit{slope} between 100 and 1000 pc is approximately the same from one sample generation to the next. The change in overall normalization is due to the Poisson noise associated with the number of planets per star, drawn in random uniform fashion between 1--3 (non-STIPs) and 5--7 (STIPs). In samples where more STIPs happen to have 7 planets, as opposed to 5, the overall planet occurrence normalization is higher. 

We note agreement overall with measured occurrence rates of planets orbiting M dwarfs: this is reasonable, given that the prescription we employed for generating planetary systems is itself derived from the \textit{Kepler} M dwarfs \citep{Ballard16}. Across all stars in the synthetic sample, the resulting true occurrence rate is 2.6$\pm0.2$ planets star$^{-1}$ (consistent with a 20\% rate of STIPs with between 5--7 planets, and 80\% non-STIP rate with 1--3 planets): the range is due to Poisson noise from the integer number of planets assigned per star. This occurrence rate is in exact alignment with the 2.5$\pm$0.2 planets star$^{-1}$ measured by \cite{Dressing15}; indeed, the example same sample was used to derive the STIP rate among M dwarfs (i.e. \citealt{Ballard16}) that we use here. Other studies have found estimates both higher and lower, but we believe that the overall normalization in planets star$^{-1}$ can vary, while the predicted trend with galactic height will manifest similarly. For example, the rate of 4.2$\pm$0.6 planets star$^{-1}$ for the \textit{Kepler} M dwarfs determined by \cite{Hsu20} using approximate Bayesian computation is substantially higher (likely due to the more sophisticated occurrence rate machinery they employ, described in more detail in the next section). \cite{Hardegree19} found a rate of 1.19$^{+0.70}_{-0.49}$ planets star$^{-1}$ for mid-M dwarfs, but only included planets with orbital periods interior to 10 days. With different assumptions about the power-law nature of the occurrence rate, \cite{Bryson20} found the rate to be even lower: 0.8$^{+0.27}_{-0.23}$ planets star$^{-1}$. Among our 25 iterations, we find high and low values of 2.9 and 2.1 planets star$^{-1}$. Because we found the \textit{slope} of the occurrence between 100-1000 pc to be similar regardless of overall normalization, the trend we explore here is more sensitive to STIP rate than to the inferred normalization in planets star$^{-1}$. 

\subsection{Recovering Injected Occurrence from Synthetic Observations}
\label{sec:recover_injected}

\begin{figure*}
    \centering
    \includegraphics[trim={0.5cm 0 0.5cm 0}, width=6.0in]{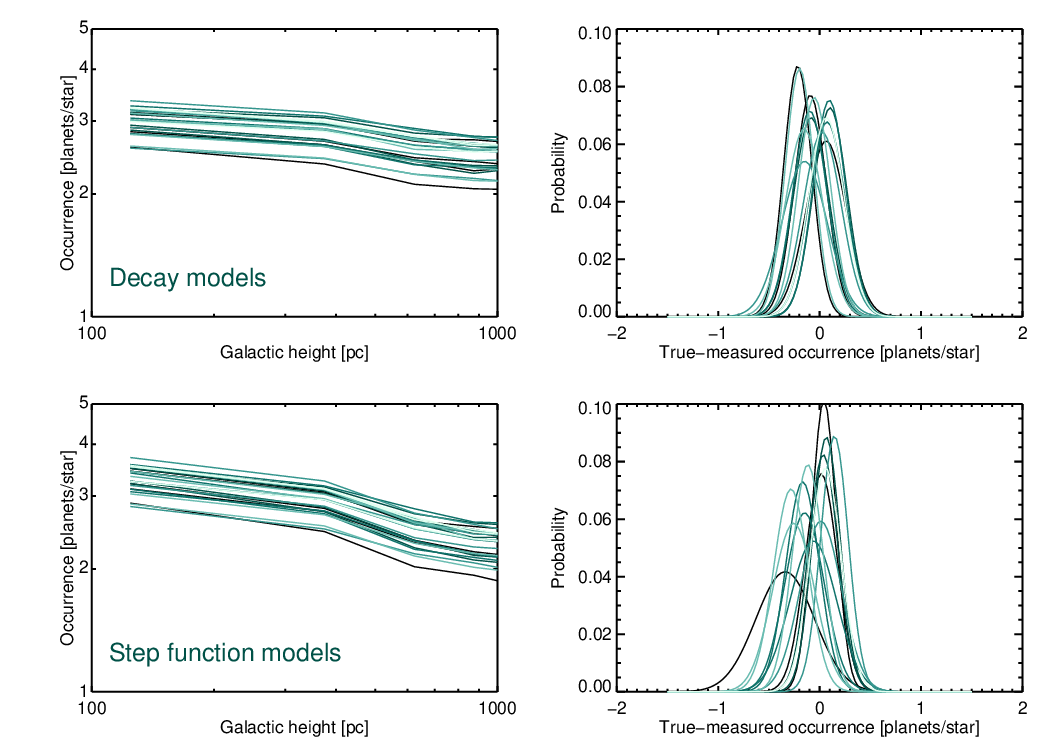}
    \caption{\textit{Left panels}: Generations of planet occurrence with galactic height between 100-1000 pc, when imposing the trends prescribed from the ``decay" (top) and "step" (bottom) model for varying STIP rate. At right, the uncertainty on the rate of planet occurrence across the entire synthetic sample, when reverse-engineering from their transit yield.}
    \label{fig:all_sims}
\end{figure*}

Once we have a sample of ``detected" planets in hand, we go on to determine a ``measured" occurrence rate from the detected sample, as in the real occurrence rate studies cited above. We follow \cite{Dressing15} in their use of the the inverse detection efficiency method (IDEM) to extract the underlying occurrence (see also \citealt{Howard11a}, \citealt{Petigura13a}, \citealt{Foreman_Mackey_2014}). With this metric, within a given cell of radius and period space, each set of \textit{detected} planets in that cell is ``upweighted" by the mission sensitivity and the transit probability to infer the true number of planets. \cite{Hsu18} has shown that IDEM underestimates the true planet occurrence rate for real detections, where the signal-to-noise is marginal. This is attributable to the use of the the \textit{estimated} planet size to compute the completeness correction, rather than the true planet size. In our simplified control experiment, we employ the true planet radius (recorded for each ``detected" planet) when reverse-engineering the occurrence. Using the true radius, as opposed to the ``measured" planet radius, we expect to avoid the bias from IDEM studies of real planets. We show in the right panel of Figure \ref{fig:all_sims} our recovered planet occurrence (across the entire stellar sample, inclusive of all galactic heights) in each of the 25 iterations, as compared to the injected planet occurrence; the distributions of true-measured planet occurrence cluster as expected around zero. 

\subsection{``Measuring" the occurrence rate with galactic height}

\begin{figure*}
    \centering
    \textbf{Selection of Simulations Employing ``Decay" of Compact Multiples}\par\medskip
    \includegraphics[trim={0.5cm 0 0.5cm 0}, width=6.0in]{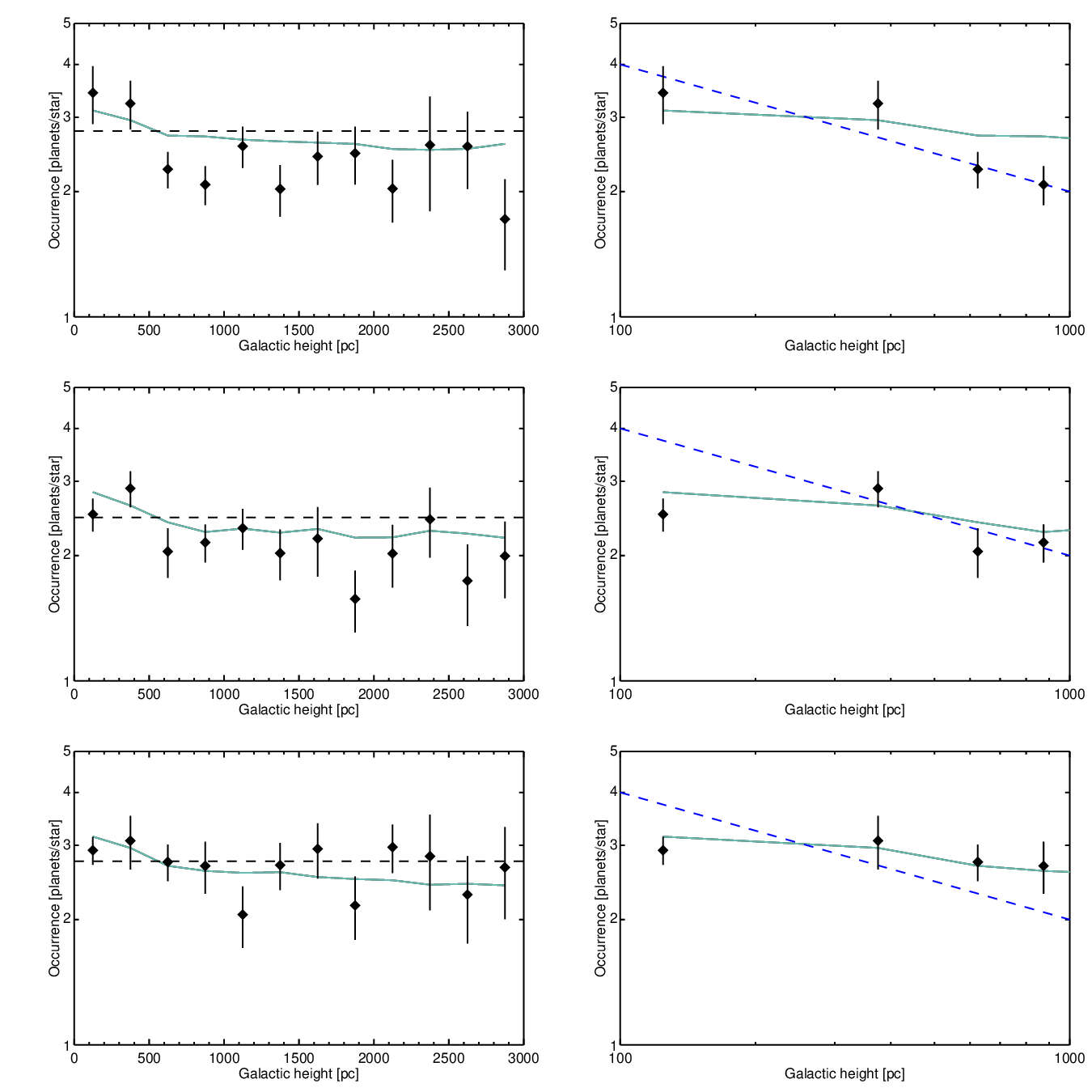}
    \caption{Three simulations of planet occurrence versus galactic height, employing a ``decay" model for compact multiple systems. \textit{Left:} Planet occurrence versus height, linearly from 0--3000 pc above the midplane. Green depicts the injected underlying population model, while black depicts the recovered occurrence for that height bin. \textit{Right:} The same model and synthetic observations, over a 1000 pc baseline and compared to the \cite{zink_scaling_2023} height trend.}
    \label{fig:decay_iterations}
\end{figure*}

\begin{figure*}
    \centering
    \textbf{Selection of Simulations Employing Step Function in Occurrence of Compact Multiples}\par\medskip
    \includegraphics[trim={0.5cm 0 0.5cm 0}, width=6.0in]{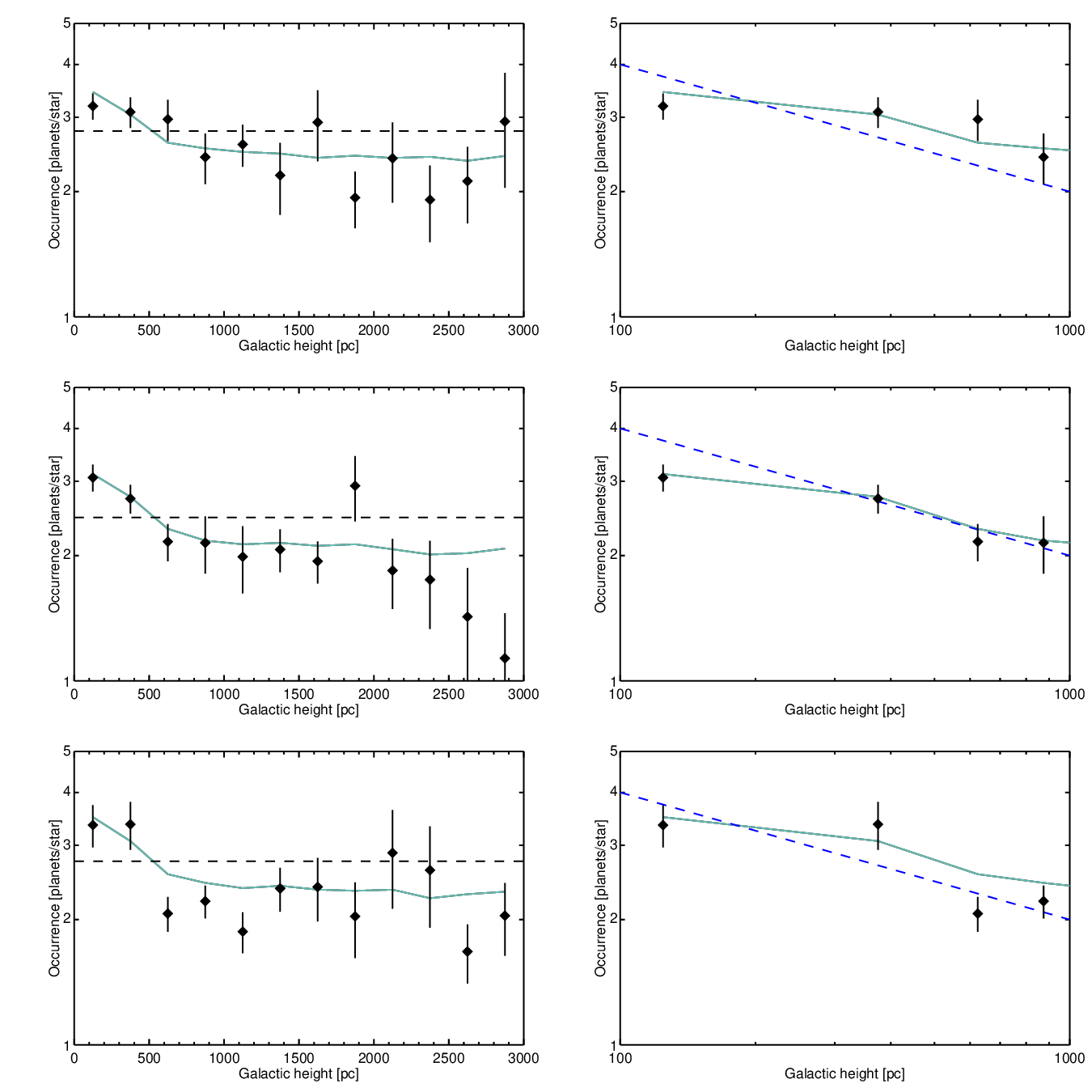}
    \caption{Three simulations of planet occurrence versus galactic height, employing a step function in galactic time for the rate of compact multiple systems. \textit{Left:} Planet occurrence versus height, linearly from 0--3000 pc above the midplane. Green depicts the injected underlying population model, while black depicts the recovered occurrence for that height bin. \textit{Right:} The same model and synthetic observations, over a 1000 pc baseline and compared to the \cite{zink_scaling_2023} height trend.}
    \label{fig:step_iterations}
\end{figure*}

In Figures \ref{fig:decay_iterations} and \ref{fig:step_iterations} we show the results of 3 random generations of stellar and planet samples, for  the ``decay" and then the ``step" models, respectively. For each generation, we divide the sample into bins of 250 pc. For each bin in galactic height, we identify the sample of transiting planets, from which we reverse-engineer the underlying planet occurrence rate for stars in that bin. We show here both the true occurrence (shown in green) and the measured occurrence (in black points, with corresponding error bars). As expected per Figure \ref{fig:all_sims}, the scatter around the ``true" occurrence rate (shown in green) is Gaussian in nature, and with typical error bars in each bin of 10-30\% (approximately the same precision as in \citealt{zink_scaling_2023}, by design from sample size selection per Section \ref{sec:sample_size}). The leftmost panels show the entire predicted trend between 0 and 3000 pc from the midplane; the occurrence rate error bars grow increasingly large as the sample size dwindles at higher galactic height. The rightmost panels show the inset from 100-1000 pc, with the \cite{zink_scaling_2023} power-law overplotted for the sake of comparison. We have adjusted the normalization upward to account for the increased planet occurrence around M dwarfs. 

We find that both of our model scenarios produce trends in which planet occurrence decreases with height from the galactic midplane, though the trend is not a continuous power law. Both models produce a ``knee" in planet occurrence between 500-1000 pc above the midplane. However, given the size of a typical error bar in planet occurrence for each slice of galactic height, there would be no robust means of distinguishing between a broken and continuous power law over the 100-1000 pc range. 

\subsection{Comparison of trend between STIP rate models}
\label{sec:compare_to_models}

For each of 25 generations of stellar/planetary samples, we repeat the exercise in generating ``measurements" of planet occurrence with height from the galactic midplane. Following \cite{zink_scaling_2023}, we elect to fit a line in log(galactic amplitude) versus log(planets/100 stars). They offer the caveat that the true underlying model may not actually resemble only a single power law-- indeed, we observe that our modeled trend exhibits a break at $\sim$1 kpc, though it is approximately log linear between 100-1000 pc. We repeat the exercise here: for each generation, we fit a linear trend to log(galactic height) versus log(planet occurrence) for 4 points between 100 and 1000 pc, applying the ``measurement errors" to a least-squares calculation of the big-fit line. We record each combination of $\{$slope, intercept$\}$ best-fit pair, for each of the 25 stellar and planetary samples, for both the ``decay" and the ``step" models. We refer to the slope of this trend as $\tau$ per \cite{zink_scaling_2023}. 

\begin{figure*}
    \centering
    \includegraphics[trim={0.5cm 0 0.5cm 0}, width=6.0in]{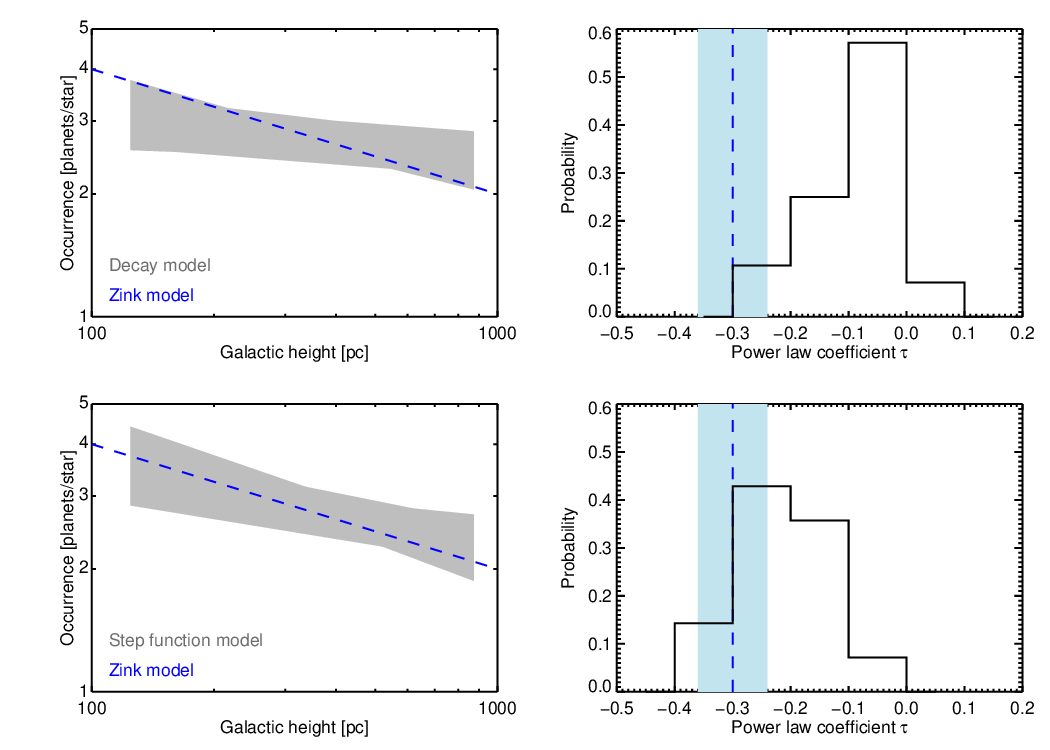}
    \caption{\textit{Left panels}: The best-fit power laws in occurrence versus galactic height, derived from  synthetic measurements of planet occurrence between 100-1000 pc described in Section \ref{sec:recover_injected}. Grey region depicts the range of slopes across all 25 iterations of the stellar and planetary sample, while blue depicts the trend from \cite{zink_scaling_2023} (normalized to higher intercept). \textit{Right panels}: Best-fit slopes $\tau$ for corresponding power laws, compared to the value and confidence interval measured by \cite{zink_scaling_2023}.}
    \label{fig:compare_to_zink}
\end{figure*}

We show the results of our findings in Figure \ref{fig:compare_to_zink}. The leftmost panels show the range of power laws associated with the best fit to the ``observations" between 100-1000 pc, across 25 iterations. This range can then be compared to the \cite{zink_scaling_2023} trend for FGK dwarfs in blue. At right, we show a histogram of the best-fit values for $\tau$ (that is, the slope) for each of the 25 samples: one with a ``decay" prescription for STIPs, and one with a ``step" prescription for STIPs. We overplot in blue the best-fit value and confidence interval for $\tau$ reported in \cite{zink_scaling_2023} of -0.30$\pm$0.06. The scatter in our $\tau$ histograms is representative of our ``measurement" uncertainty in $\tau$. We report from our ``decay" model predictions $\tau=-0.09\pm0.07$, while we find $\tau=-0.21\pm0.08$ for the ``step" model predictions. The similarity in the confidence interval is expected, due to the sample size selection described in Section \ref{sec:stellar}. It is of interest to compare to \cite{zink_scaling_2023}, though we consider different stellar populations here. A``decay" model in which STIP systems are metastable at birth, self-exciting over $\sim$Gyr timescales, results in a substantially more shallow trend in occurrence with galactic height to match.  A step function model in galactic time mimics the observed slope among FGK dwarfs more closely. That is, if the likelihood of a given star forming a STIP increased in the past several Gyr by a factor of several, the expected trend in planet occurrence with galactic height approximates the \cite{zink_scaling_2023} trend. 

\section{Results \& discussion}
\label{sec:results}

We investigate the hypothesis that a change in the rate of Systems of Tightly-packed Inner Planets (STIPs), over the course of the galaxy's planet formation history, can mimic a decrease in planet occurrence with galactic height. Focusing upon M dwarfs, we consider two different prescriptions for how the STIP rate varies, described in Section \ref{subsection:decay}. In this Section, we investigate physical mechanisms that could plausibly produce the modeled trend in STIP rate. We limit here our considerations to mechanisms that have some existing observational link to the occurrence of STIPs: stellar metallicity, Jovian companions, and stellar clustering at birth/binarity (of course, these properties are themselves related and cannot be considered only in isolation). We comment on whether existing relationships between these conditions and STIP occurrence are consistent with an increase in STIP rate over the past few Gyr of galactic history.  

\subsection{Metallicity considerations}
\label{sec:metallicity}

One of the the intringuing features of the \cite{zink_scaling_2023} finding is that the increase in planet occurrence toward the galactic midplane cannot be explained by the galactic metallicity gradient. This is consistent with the \cite{yang_planets_2023} finding that the number of planets per star increases around younger stars, even controlling for metallicity. The phenomenon by which average stellar metallicity decreases away from the midplane, has been increasingly closely studied with generations of large stellar surveys (see e.g. \citealt{mayor_chemical_1976, andrievsky_using_2002, magrini_evolution_2009, luck_distribution_2011, bergemann_gaia-eso_2014, xiang_evolution_2015,yan_chemical_2019, hawkins_chemical_2023}). The link between individual stellar metallicity and planet occurrence has been similarly studied in increasing detail. It is typically viewed as a deterministic relationship, in which the relative enrichment of metal content in the primordial disk directly affects both planet sizes and orbital properties (\citealt{santos_metal-rich_2001, Fischer_2005, schlaufman_kepler_2011,Johnson10, Buchave12, Dawson13, wang_planet_2015, Brewer18, Petigura18, anderson_higher_2021, boley_first_2024, Lu20} among many others; see also review by \cite{adibekyan_star-planet_2014}). 

In this study we have considered specifically the STIP rate, which some studies have found to increase around metal-poor host stars (\citealt{Brewer18} for FGK dwarfs, \citealt{Anderson21} for M dwarfs), while others have found no significant difference in metal content between hosts to systems of multiple transiting planets and hosts to singly-transiting planets \citep{romero_no_2018, weiss_california-kepler_2018, zhu_influence_2019}. In any case, a relative metal-poorness of compact multiples would produce the opposite effect with increasing galactic height: we hypothesize here that the STIP rate \textit{increases} toward the galactic plane, and metallicity is increasing, rather than decreasing, in this direction on average. This metallicity preference around metal-poor stars for STIPs, if it exists, cannot explain a strong increase in STIPs over the past several Gyr. On the other hand, \cite{Lu20} found that the small-planet occurrence–host star metallicity relation is even stronger for M dwarfs than for solar-type stars: in this sense, metallicity trends may trace trends with galactic height more closely among M dwarfs.  

\subsection{Varying rate of companion Jupiters}
\label{sec:jupiters}

We also consider the possibility of a link between the STIP likelihood and the existence of companion Jovian planets. While once thought to be extremely rare around small stars \citep{Johnson12}, more recent studies of early-type M dwarfs observed with \textit{TESS} found that they host Hot Jupiters at rates only a factor of a few less often than Sunlike stars (\citealt{gan_occurrence_2022}, see also \citealt{bryant_occurrence_2023}). With respect to further-out giant companions, they are less common around M dwarfs (particularly late-type M dwarfs, per \citealt{pass_mid--late_2023}), but still occur at $\sim$25\% their rate around FGK dwarfs \citep{Clanton16}. 

The hypothesized relationship could take several forms. Some studies demonstrate that  emergence of cold Jupiters at can drive dynamical instabilities in systems that are closed packed \citep{matsumura_effects_2013, Huang17, Lai17,pu_eccentricities_2018, Becker17}. In this sense, STIPs would reside preferentially around stars without such Jupiters. Other investigations have concluded that the formation of STIPs (that is, systems of small close-in planets) ought to be hampered early on by the presence of a Jupiter. The resulting anti-correlation between cold Jupiters and small inner planets would either result from (1) preventing the inward migration of such planets after their formation \citep{izidoro_gas_2015} or by considerably reducing or even halting the inward flux of pebbles to form such planets \citep{lambrechts_formation_2019, mulders_why_2021}. Yet, a \textit{positive} correlation between cold Jupiters and systems of small inner planets was then observed by \cite{Zhu18} as well as \cite{bryan_excess_2019}; though other studies did not observe the correlation \citep{bonomo_cold_2023}. More recently, \cite{bryan_friends_2024} identified that the positive relationship between Jupiters and inner small planets is metallicity dependent: it exists for metal-rich stars, but not for metal-poor stars (though this study excluded M dwarfs from consideration). 

Importantly for this consideration, \cite{miyazaki_evidence_2023} identified that Hot Jupiters (with orbital periods 1–10 days) occur more often around young stars, while cold Jupiters (with periods of 1–10 yr) show no such preference. The change occurs with a timescale of order Gyrs. If we aim to reconcile this finding with the hypothesized increase in the STIP rate over the same approximate timescale, one possible scenario is that the formation of hot-Jupiters is an extreme outcome of early metastability of STIPs \citep{boley_situ_2016}. In this sense, the increasingly STIP rate would conceivably drive a corresponding increase in Hot Jupiters. One of the  \cite{miyazaki_evidence_2023} models for Hot Jupiter occurrence, supposed by their observations, takes the form of an exponential increase over Gyr. It results in an increase in the occurrence of Hot Jupiters by a factor of $\sim$5 within the past 2 Gyr. If indeed the STIP rate is rising, and a common fraction of these STIPs result in Hot Jupiters (via the proposed in-situ formation scenario) then these quantities would tend in a common direction. 

Given that cold Jupiters are associated with increased likelihood of small inner planets (at least for metal-rich stars, per \citealt{bryan_friends_2024}), a recent increase in cold Jupiters would be an appealing story toward a recent higher STIP rate. Indeed, \cite{chachan_small_2023} predict that as long as colder Jupiters exist, their correlation with the inner super-Earths remains and potentially enhances even around M dwarfs. However, the fact that cold Jupiters (with periods of 1–10 yr) do not show such an increase over the past few Gyr \citep{miyazaki_evidence_2023}, at least around FGK dwarfs, suggests they are not an obvious candidate to explain a recent steep STIP increase.   

\subsection{Varying stellar birth environment}

The traditional picture of star-planet formation occurs in relative isolation \citep{armitage_dynamics_2011, williams_protoplanetary_2011, winn_occurrence_2015}. Under isolated conditions, the stellar and disk properties alone are presumably responsible for the planetary outcome. This is why the metallicity of the host star, presumably reflective of the composition of the protoplanetary disk, has historically been assume to have a deterministic relationship to planet formation (see Section \ref{sec:metallicity} above). However, recent findings present the case that the stellar-metallicity-to-planet relationship may be partly or wholly a proxy for other properties (such as age), which themselves trace metallicity  \citep{winter_prevalent_2020, kruijssen_bridging_2020, miyazaki_evidence_2023}. A strongly deterministic component of planet formation, within this framework, may be the stellar density of the nearby environment and the host star's dynamical history \citep{adibekyan_stellar_2021, dai_planet_2021, winter_planet_2024, cai_signatures_2018, kruijssen_bridging_2020}. If these properties are changing over galactic time, they may simply covary with metallicity, without the metallicity \textit{per se} being causally linked to planet formation. More probably, planetary outcomes are determined by a mixture of properties of the host star and its environment, with that mixture potentially varying with galactic environment. In this sense, planet occurrence may vary across ``chemo-kinematic" space, in a way newly traceable by \textit{Gaia} \citep{veyette_chemo-kinematic_2018, carrillo_know_2020}. 

In plausibly linking stellar birth environment with STIP rate, it is useful to consider the \cite{longmore_impact_2021} result: they found that the planet multiplicity distribution varies, dependent upon whether the host star resides in a high or low stellar phase space density environment. There is a greater fraction of multiplanet systems around field stars than in overdensities. A recently higher STIP rate might therefore plausibly emerge from planet formation in lower density stellar environments. An investigation of this possible link would require careful consideration of the relationship between the stellar density of the formation environment, and the rate of binarity, itself related to the architectures of planetary systems \citep{sullivan_revising_2023}. Recent findings from \textit{Gaia} illustrate that such environmental trends can be traced over galactic time. For example, \cite{niu_binary_2021} found that the binary fraction of G\&K dwarfs is lower in the thin disk than the thick disk. If M dwarf binary fraction is similarly lower in recent galactic history, it could indicate a difference in stellar natal environment that might also imprint upon planetary systems.

\subsection{Sensitivity to model parameters}
We have assumed a single fiducial model for both the ``step" and ``decay" trends with STIP occurrence. Tuning these models will correspondingly shift the resulting trend in galactic height. In particular, it is useful to consider whether there exists an alternate ``decay" function that could be tuned to produce a steeper slope; as discussed in Section \ref{sec:compare_to_models}, the given decay law presents a slope barely distinguishable from 0 with the given sample size. However, \cite{lam_ages_2024} considered various forms of this functional decay law: the one we employed here (by which sculpting potentially continues over 10 Gyr) is considered the ``most" sculpting that is consistent with the \textit{Kepler} yield. Selecting an even steeper ``decay" over longer timescales results in a fraction of STIPs too low to match the transit multiplicity yield from \textit{Kepler}. We conclude that if we made the decay law ``stronger" in service of matching the \cite{zink_scaling_2023} (that is, if close to 100\% of STIPs were sculpted into non-STIPs over a period of several Gyr), the result would no longer reproduce the observed \textit{Kepler} sample of transit multiplicity (seen the rightmost panel in Figure \ref{fig:stip_models}). There do exist decay models that produce ``less" sculpting, but these would furnish an even shallower slope with age (and correspondingly, galactic height). 

Similarly, the shape of the ``step" function by which STIP occurrence increased in recent galactic history could be tuned, with changes to the resulting slope in planet occurrence with galactic height. Here we are again constrained by the need to reproduce the present-day STIP rate of $\sim$20\%. For example, we could similarly induce the STIP fraction to ``step" from 8\% for the first 10 Gyr of galactic history, to a rate of 50\% in the past 2 Gyr; this would result in the same correct STIP rate today, but produce too \textit{steep} of a slope to match the \cite{zink_scaling_2023} finding. The step function model is also highly simplified: a possibility for future work is the effect of a slower linear increase in STIPs extending over a longer period of time.

\section{Conclusions}
 We have investigated how a varying rate of Systems of Tightly-Packed Inner Planets (STIPs) over Gyr could appear as a trend in planet occurrence with galactic height. We have confined this study to the predicted effect around around M dwarfs, given their high rate of STIPs and their usefulness as probes of galactic kinematics. The recent findings of \cite{zink_scaling_2023}, in which the rate of small planets increases toward the galactic midplane for FGK dwarfs, points to an increasingly wide scope for occurrence studies. With the \textit{Gaia} data releases, investigations into the galactic context of planet occurrence are newly possible \citep{winter_planet_2024, longmore_impact_2021, kruijssen_bridging_2020, adibekyan_stellar_2021, bashi_small_2019, dai_planet_2021}. 

We focus our consideration upon two fiducial models by which STIP fraction varies with time. Both models produce an apparent increase in planet occurrence around younger stars, which are statistically closer to the galactic midplane. By applying these prescriptions to a representative synthetic sample of stars, with galactic height assigned from age, we investigate how the resulting trend would manifest. We find that that a recent ($\sim$2 Gyr ago) increase in the STIP rate by a factor of several produces a similar slope to the trend observed by \cite{zink_scaling_2023}.  This ``step" model that we set forth in Section \ref{subsection:decay}, involves an increase in the rate of STIPs formed at birth, from a baseline of 10\%  for the first 10 Gyr of the Milky Way's history, to a rate of 50\% in the most recent 2 Gyr. It results in an approximate factor of 2 decrease in the average number of planets star$^{-1}$, comparing host stars at 100 versus 1000 pc from the galactic midplane. In contrast, the ``decay" model involves a single mode of planet formation and evolution over galactic time. Systems become dynamically hotter with age, a mechanism which is physically plausible \citep{Pu15, zinzi_anti-correlation_2017} and supported by some observational evidence \citep{yang_planets_2023}. It results in a much shallower decrease in planet occurrence with increasing galactic height, changing the number of planets star$^{-1}$ by a factor of $\sim$1.2 between 100-1000 pc. Importantly, both models replicate the overall 20\% STIP rate observed among current M dwarfs \citep{Ballard16, Muirhead15}. 

We intend the investigation here to serve as a proof-of-concept, but we have made many simplifying assumptions. A more complex modeling endeavor might include other functional forms for a proposed increase in STIP occurrence, and investigate the effect of varying the age distribution of the stellar sample. We have approached the problem employing a sample of M dwarfs, which allows us to make the simplifying assumption that no stars leave the Main Sequence during Milky Way history; a sample employing FGK dwarfs could be compared more directly to the \cite{zink_scaling_2023} study. We have forward-modeled, as our synthetic observable, the number of planets per star; this is also the metric employed by \cite{zink_scaling_2023}. We have not varied as part of this study the \textit{fraction of stars with planets}; this is a separate quantity and scales differently with metallicity \citep{zhu_influence_2019}. Future studies varying that fraction, perhaps jointing with the number of stars per planet, could identify how those quantities would need to combine to mimic the observed occurrence trend.


\bibliography{main.bib}

\end{document}